\documentclass[aps,twocolumn,secnumarabic,nobalancelastpage,amsmath,amssymb,
nofootinbib]{revtex4}

\usepackage{xcolor}
\usepackage{hyperref}
\hypersetup{breaklinks=true, hidelinks}
\usepackage{harmony} 
\usepackage{float}
\usepackage{graphics}     
\usepackage{graphicx}      
\usepackage{longtable}     
\usepackage{url}          
\usepackage{bm}            
\usepackage[utf8]{inputenc}
\usepackage[spanish]{babel}

\DeclareMathOperator*{\uplusop}{\uplus}

\begin{document}

\title			{Polyrhythmic Harmonies from the Sky: Transforming Satellite Images of Clouds into Musical Compositions through Algorithms}
\author         {Carlos Darío Badilla Cerdas\footnote{\url{carlos.badillacerdas@ucr.ac.cr} $|$ \url{dari.badilla@gmail.com}}}
\affiliation    {Escuela de Física, Universidad de Costa Rica}

\begin{abstract}
In a context of increasing scientific specialization and deficiencies in the scientific literacy of the population, there arises a need to broaden the methods of scientific dissemination. This study proposes an approach that combines music with scientific concepts, focusing on the sonification of satellite images as the core. A generative musical composition system is developed that uses visual data to create accessible and emotional auditory experiences, thus enriching the fields of scientific dissemination and artistic expression. It concludes with an example of the algorithm's use in a musical composition.
\end{abstract}

\maketitle

\section{Introduction}
Inspiration can come from anywhere, from the music we listen to daily to observing phenomena as common as clouds. Since the dawn of humanity, clouds have been a source of inspiration for countless people: from those seeking recognizable shapes in their serene abstraction to modern scientists trying to predict their behaviors through complex mathematical models.

The inspiration for this work stems from the fusion of the dynamic and ephemeral beauty of clouds with Stratus, a musical instrument created by composer Ólafur Arnalds \cite{stratus, stratus2}. The idea of developing an algorithm that integrates clouds into sonification has been on my mind for several years. Over time, I acquired the necessary knowledge to materialize this idea as part of an artistic profile aimed at sparking curiosity about science through scientific data sonifications.

With the growing and unstoppable specialization of science, the gap between the general population and the scientific community has widened significantly. In an era where technology allows us to have all human knowledge at our fingertips, scientific literacy remains surprisingly deficient. In this context, following the legacy of Thomas Henry Huxley \cite{schwsartz1999robert}, modern science communicators have emerged. Using mass media such as social networks, these communicators attempt to convey scientific concepts to millions of people around the world. However, they face numerous challenges, from the difficulty of simplifying complex concepts to the limitations imposed by the formats and rules of each platform, which often restrict clarity and depth of content.

Moreover, it is crucial to consider the public’s attention span and the shifting trends in digital media, which further complicates maintaining the relevance of scientific dissemination content. In this context, expanding the domains of the field can be beneficial, combining scientific concepts with other disciplines, such as performing or musical arts, with the goal of capturing the general public's attention towards the world of science.

This research is based on the principles mentioned above, seeking a method to combine music with striking scientific concepts. Specifically, it focuses on space sciences using satellite imagery as the core. This approach aims not only to be part of an audiovisual medium for dissemination but also to serve as an effective tool for artistic expression.

In the realm of space science dissemination, sonification holds immense potential. The vastness and complexity of outer space often overwhelm the general public, making it difficult to understand and appreciate astronomical phenomena. Music based on space phenomena offers an engaging way to bring these concepts closer to the audience, transforming visual data into accessible and emotional auditory experiences. This transformation not only facilitates understanding of the data but can also generate greater interest and enthusiasm for space sciences.

Renowned institutions, such as NASA, have begun to implement sonifications for their content dissemination. A notable example is the work of the Chandra X-Ray Center \cite{Chandra}, which has developed sonifications of celestial objects to capture the general public’s attention. These are accompanied by videos illustrating the process undertaken to create the sonification, thereby facilitating a better understanding of the method and its results.

\section{Sonification}
Sonification, in its most basic conceptualization, is explained as the representation of data as sound, according to The Sonification Handbook \cite{hermann2011sonification}. This process converts data into auditory patterns, opening new ways of understanding and perceiving information. Examples that employ this idea are abundant, from the simple tick of the Geiger counter \cite{korff2013geiger} to the complexity of the auditory interpretation of Saturn’s magnetic field shape captured by the Cassini probe \cite{nasa}.

There are three main types of sonifications: parameter-based, model-based, and direct data translation (audification) \cite{ballora2014sonification}. To determine the type of sonification that best suits each phenomenon, it is necessary to analyze the properties of the system and the specific objective of the sonification. In addition to these three types, Gresham-Lancaster \cite{gresham2012relationships} proposes two additional classifications based on their purpose: first-order sonifications and second-order sonifications. First-order sonifications aim for a direct link between the data used and the resulting sound, while second-order sonifications take data sets and, by applying the corresponding algorithmic processes, generate a sound that does not necessarily serve as a direct link to the phenomenon used.

In this context, the present research aims to create a second-order sonification using algorithmic processes on the input images. A final relevant classification is the differentiation between sonification and musicalization. Musicalization is the result of applying sonification following the rules of a specific musical system. Thus, all musicalization is a sonification, but not all sonification is a musicalization. Consequently, the result of this research is a second-order musicalization based on a parametric model.

Since every phenomenon in the universe generates information \cite{burgin2022information}, any phenomenon can be sonified. Sonification encompasses any area of knowledge, whether with research capabilities (first-order sonification) or artistic capabilities (second-order sonification). Various examples of sonification in our times include the aforementioned case of NASA, but there are many other topics, such as the sonification of meteorological data \cite{kalonaris2022tokyo, arai2012sonification, polli2005atmospherics, childs2003using, woo2023weatherchimes, george2017making}, gravitational waves \cite{Ligo}, galactic catalogs \cite{bardelli2021}, and social behaviors \cite{Carlos}, among others. Some of these efforts are driven purely by artistic creation and its complement with the scientific background, as in the case of Interactive Sonification of Weather Data for The Locust Wrath, a Multimedia Dance Performance \cite{lindborg2018interactive}, where an interactive dance performance with live sonification is presented, which sonifies climate data from Southeast Asia between 1961 and 2000. Cases like this last one drive the creation of projects like this research, where sonification is used as an artistic tool with dissemination purposes and not for academic research purposes.

\section{Data}
\begin{figure}
\centering
\includegraphics[width= 8 cm]{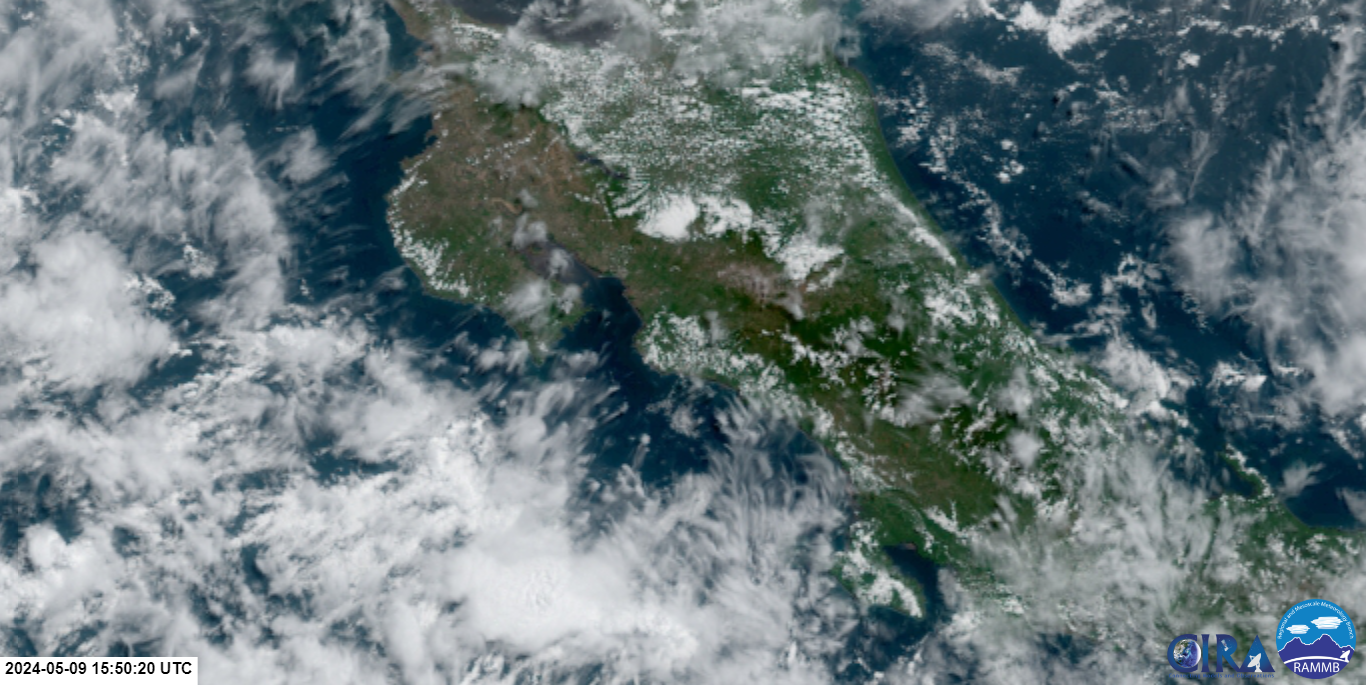}
\caption{Satellite image of Costa Rica obtained from the RAMMB/CIRA SLIDER: GOES-16 portal, on 05/09/2024 around 10:00 local time.}
\label{sat}
\end{figure}

For the musicalization, satellite image sequences from the GOES-16 satellite were used, obtained through the RAMMB/CIRA SLIDER: GOES-16 portal \cite{rammb_slider}. This portal offers a wide range of functionalities, allowing users to specify precise geographic coordinates, define the time lapse between images, and apply various filters that alter the visual characteristics of the final images. It also allows downloading the data in various formats. In the context of this study, images in .png format were selected, $\textit{frame per frame}$, in the visible wavelength, centered over the territory of Costa Rica with a resolution of $~50 Km$. The images were captured at intervals of one hour. Figure \ref{sat} shows an image belonging to the data used in this project.

\section{Sonification Proposal}
The aim of this research is to develop a generative musical composition system that uses satellite images as its core. The components and the functioning of the proposed system are described below.

The system consists of a set of satellite images captured with a one-hour interval between each image. Based on the principle of wind chimes, which generate sound when moved by the wind, a system has been conceived that activates simple rhythmic patterns based on cloud density. Clouds, like wind chimes, are carried by the wind.

\begin{figure}
    \centering
    \includegraphics[width= 8 cm]{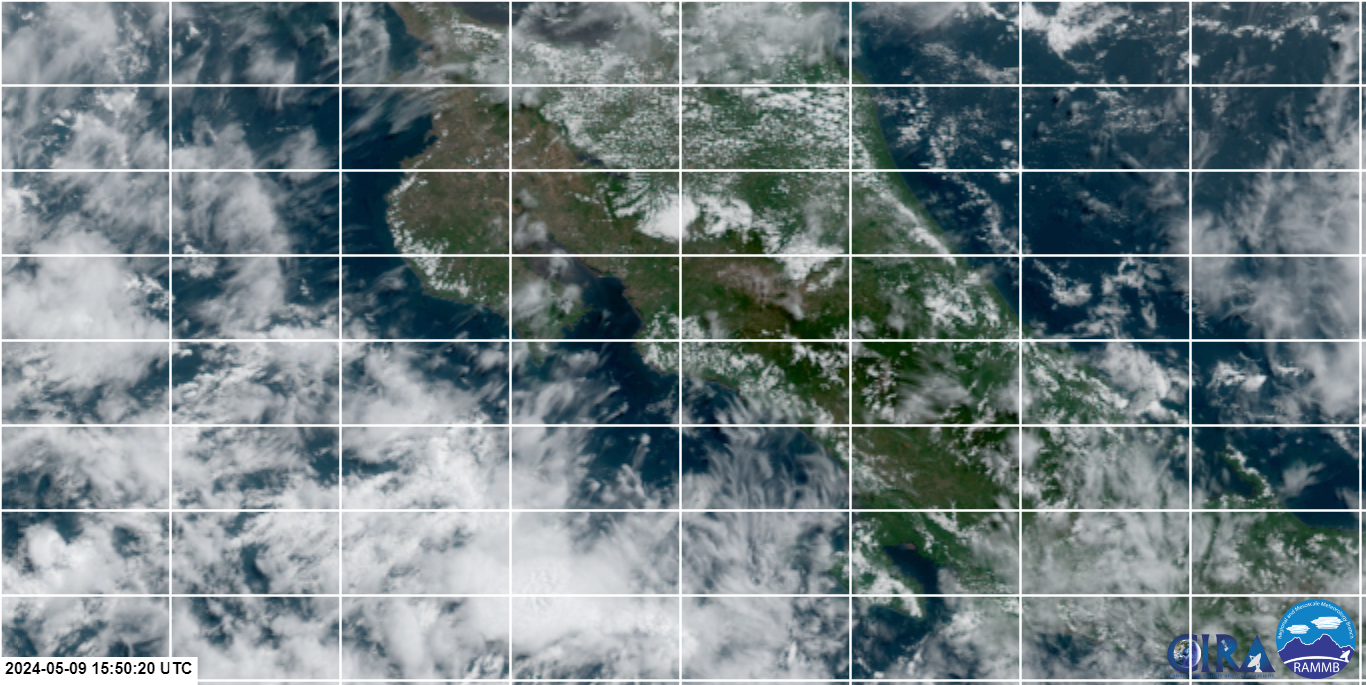}
    \caption{Satellite image of Costa Rica divided into 64 regions.}
    \label{sat_div}
\end{figure}

The designed algorithm analyzes each image in the sequence individually. To do this, each image is divided into $n^{2}$ equal regions, and the cloud density in each quadrant is evaluated. If the density exceeds a predefined threshold, the musical pattern corresponding to that quadrant is activated. This process is repeated for each image in the sequence. The albedo of the clouds is used for this, with higher reflection of incident light indicating higher cloud concentration. Figure \ref{sat_div} presents an example of the subdivision of images into regions.

As a compositional tool, the system allows the composer to define the specific musical function for each quadrant of the image. For example, if an image is divided into 8 columns and 8 rows, the composer must define 64 individual patterns.

The objective is to generate complex patterns from simple elements. Each quadrant of the image contains a basic musical pattern, consisting of a single note repeated five times over a certain period of time. For example, a pattern could consist of five quarter notes with the note $C_{4}$. The note, starting time, and temporal pattern can be unique for each quadrant. The combination of multiple quadrants activated simultaneously will result in complex polyrhythmic patterns that could not be achieved otherwise.

To add greater dynamism to the final work, a parameter is introduced that varies the intensity of the musical note in a wave-like manner. The first note begins with low intensity, gradually increasing to its maximum, and then gradually decreasing until it is silent. The maximum intensity is proportional to the cloud density of the quadrant. Thus, just as clouds arrive in waves, the musicalization maintains the pattern, reinforcing the initial analogy.

\begin{figure}
    \centering
    \includegraphics[width= 8 cm]{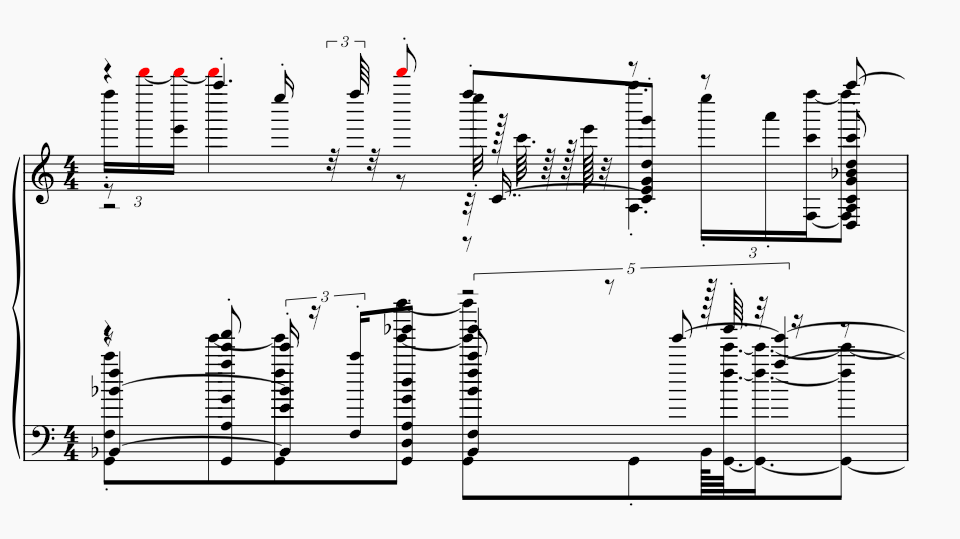}
    \caption{Random measure in the final score.}
    \label{partitura}
\end{figure}

Given the complexity of the proposed patterns, the musicalization is specifically designed to be performed by a computer. Figure \ref{partitura} shows a single random measure from the final score for piano.

The final result of the musicalization will largely depend on the patterns defined by the composer and the cloud density in the analyzed satellite images. 

\section{Mathematical Formulation of Sonification}
Since a musical score is an arrangement of objects along the staff, i.e., a rectangular arrangement, it can be interpreted as a matrix. By doing this, it is possible to mathematically establish the algorithm.

We propose a matrix $\mathbb{P}_{n \times nm}$ such that:

\begin{equation}
    \mathbb{P}_{n \times nm}  \uplusop_{p=1}^{H} t_{cp} \cdot \left( \mathbb{M} \odot \mathbb{S}_{p} \right),
\end{equation}

where $\odot$ represents the Hadamard product and $c$ the number of measures that each image covers in the composition. In this case, the operator $\uplusop$ is defined as the iterative concatenation of matrices. Each element $\mathbb{P}_{p} = t_{p} \left( \mathbb{M} \odot \mathbb{S}_{p} \right)$ is concatenated horizontally, forming the block matrix:

\begin{equation}
    \mathbb{P} = \begin{pmatrix}
        \mathbb{P}_{1} & \mathbb{P}_{2} & \cdots & \mathbb{P}_{H}
    \end{pmatrix}
\end{equation}

where each $\mathbb{P}_{p}$ is a musical block generated by a single image in the sequence. Thus, $\mathbb{P}$ is equivalent to a musical score on the staff, where each entry indicates the note and the time at which it should be played.

Once this is established, each component of the right-hand side of the equation is defined as follows:

\subsection{Musical Coefficients Matrix}

The matrix $\mathbb{M}$ corresponds to the musical coefficients, which dictate the note, the octave, the repetition, and the duration of each note.

\begin{equation}
 \mathbb{M}_{n \times n} := \begin{pmatrix}
    m_{11} & \ldots & m_{1n} \\
    \vdots & \ddots & \vdots \\
    m_{n1} & \ldots & m_{nn}
 \end{pmatrix}   
\end{equation}

where each coefficient of the matrix corresponds to a vector that stores the notes:

\begin{equation}
    m_{ij}  =  \sum_{l=1}^{m} S_{l} \cdot t_{n + D + k \cdot l} \cdot f_{l}
\end{equation}

Where $S_{l}$ is an element of the resulting set from the Cartesian product of the scale set $N$ and the octave set $\Omega$. This definition allows for easy variation of the instrument's range:

\begin{equation}
    S = \Omega \times N = \{ (\omega, \nu) \mid \omega \in \Omega, \nu \in N \}
\end{equation}

The value of $N$ corresponds to:

\begin{equation*}
    N = \{  I, II, \cdots, VII \} \subseteq \{  C = 1, C\# = 2, \cdots, B = 12 \}
\end{equation*}

While $\Omega$ corresponds to:

\begin{equation*}
    \Omega = \{ 0, 1, \cdots, 7 \}
\end{equation*}

The value of $k$ corresponds to an element of the duration set, being the value of the musical figure corresponding to the quadrant:

\begin{equation*}
    k \in \left\{ \frac{2}{3}, 1, \frac{3}{2}, \frac{4}{3}, 2, \frac{8}{3}, 3, 4, 6 \right\} = \left\{ \overset{3}{\Vier}, \Vier, \cdots, \Ganz\Pu \right\}
\end{equation*}

The value of $t_{n}$ is the current beat in the global counter $n$, and finally, $D$ corresponds to the delay applied per quadrant.

While $f_{l}$ is the intensity of the note, following the function:

\begin{equation}
    f_{l} = \left| \sin \left( \frac{2l}{\pi} \right) \right| \cdot \sum_{i,j=1} \frac{p_{ij}}{n^{2}}
\end{equation} 

where the summation term corresponds to the average of all elements (pixels) of the submatrix corresponding to each quadrant of the image.

\subsection{Activation Matrix}

The activation matrix $\mathbb{S}$, a matrix of ones and zeros, determines whether the note housed in $\mathbb{M}$ is activated:

\begin{equation}
 \mathbb{S}_{n \times n} := \begin{pmatrix}
    s_{11} & \ldots & s_{1n} \\
    \vdots & \ddots & \vdots \\
    s_{n1} & \ldots & s_{nn}
 \end{pmatrix}   
\end{equation}

where:

\begin{equation}
    s_{ij} = \chi_{A}(x)
\end{equation}

where $\chi_{A}(x)$ is the indicator function such that:

\begin{equation}
    \chi_{A}(x) = \begin{cases}
        1, & \text{if } x \in A \\
        0, & \text{if } x \notin A
    \end{cases}
\end{equation}

where $A = \{ a \in \mathbb{R} \mid a \geq b \}$, with $b$ being the desired percentage of cloud cover to exceed. The value of $x$ is obtained by:

\begin{equation}
    x =  \sum_{i,j=1} \frac{p_{ij}}{n^{2}}
\end{equation}

is the average of all elements (pixels) of the submatrix corresponding to each quadrant of the image.

Thanks to its mathematical formulation, the algorithm is invariant under the change of musical system. This allows it to be adapted to different systems, such as the use of pure frequencies or the slendro, typical of gamelan music.

\section{Results of the Algorithm}
For the present musicalization, the images were divided into 8 columns and 8 rows, resulting in a total of 64 quadrants.

The result of the final musicalization depends on the definition of the coefficients of the matrix $\mathbb{M}$, corresponding to each quadrant of the image. The definition of the parameters of each coefficient is entirely at the discretion of the composer. Functions with rhythmic patterns were defined that only depend on the repetition of the same note. For example, using the C scale, in the entry $m_{11}$, the function was defined as a repetition of 5 quarter notes with the note $\text{C}_{1}$, while for the entry $m_{38}$, the function is a pattern of triplet half notes on the note of $\text{B}_{5}$.

\begin{figure}
    \centering
    \includegraphics[width= 7.5 cm]{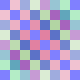}
    \caption{Final configuration of the matrix $\mathbb{M}$ used for the musicalization.}
    \label{patron1}
\end{figure}

Different configurations were tested until reaching the final one, where the melodic qualities of each were evaluated. Finally, a symmetrical configuration was chosen, i.e., $\mathbb{M}^{T} = \mathbb{M}$. In addition, a band matrix configuration was adopted, where each diagonal next to the main one corresponds to a degree of the scale, varying its rhythmic pattern and its position in the octave. Figure \ref{patron1} shows a visual example of the final configuration of the matrix $\mathbb{M}$.

This configuration was chosen due to the nature of the geography of the Costa Rican territory, on which the arrangement of images is centered. By aligning the globe with the vertical axis, the Central Mountain Range crosses the territory diagonally from northwest to southeast. This mountain range is so influential in the Costa Rican climate that its barrier effect is visible from satellites, which can prevent clouds from crossing from coast to coast. Thus, a diagonal configuration is optimal to take advantage of the geographical properties of the territory used and its influence on the clouds.

The choice of this configuration is based not only on the geographical representation but also on its ability to generate a cohesive and harmonious sound. The diagonal arrangement allows rhythmic patterns to develop organically as clouds arrive from the Caribbean, reflecting the natural transitions in cloud density observed in satellite images.

\begin{figure}
    \centering
    \includegraphics[width= 8 cm]{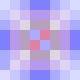}
    \caption{Example of a rejected pattern for the matrix $\mathbb{M}$.}
    \label{patron2}
\end{figure}

In contrast, other configurations were evaluated and ultimately rejected due to their lack of melodic coherence or their inability to adequately adapt to the geographical characteristics of the territory. Figure \ref{patron2} shows one of the rejected patterns for the matrix $\mathbb{M}$. This alternative configuration presented a more random arrangement of the coefficients, resulting in a disorganized and less representative musicalization of the satellite data.

\subsection{Creation of the Musical Piece}
It is worth noting that the algorithm is a compositional tool, so the choice of scale and rhythmic patterns, as well as what is considered musically pleasing, are considerations of the composer. The results will vary depending on the person using the algorithm and their artistic interpretation.

For the creation of the example in this research, the G Dorian mode was used as the scale, and the time span covered from May 8, 2024, to May 14, 2024, starting and ending at midnight each day.

The algorithm was implemented in the composition of a work belonging to the ambient genre \cite{Eno}, characterized by its ability to be ignored without creating a void or becoming the focal point in the space where it is played. The choice of this particular genre is due, besides the personal taste of the composer, to the nature of the musicalization. Since the musicalization uses the environment (the passage of clouds seen from the sky) as its core, creating a work intended to be part of the environment seems reasonable. The result is a unique musical composition, where the polyrhythmic harmonies reflect the complexities and variations of cloud formations. Because the atmosphere is a dynamic, complex, and chaotic system, the generated melodies will never repeat, maintaining the ephemeral characteristic of the intrinsic beauty of clouds.

The final work can be divided into two parts: the harmonic and the melodic. The harmonic part of the piece is performed by a cello and a synthesized pad, while the musicalization was performed by keyboard instruments like the piano, providing a chaotic melody guided by the clouds. The simplicity of the ambient genre allows the musicalization's participation to stand out, highlighting its presence without overloading the composition.

\section{Final Work}
\begin{figure}
    \centering
    \includegraphics[width= 7 cm]{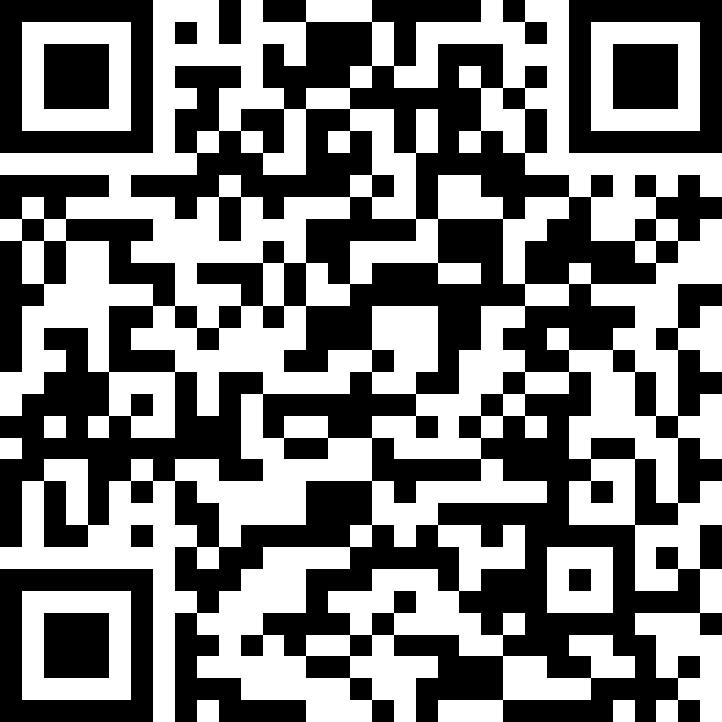}
    \caption{QR code that directly links to the final work hosted on Bandcamp.}
    \label{qr}
\end{figure}

This section presents the final result, created with the described musicalization. The work is titled \textit{This silence Made me feel empty}. The reader can access the complete work through the following link or by scanning the QR code in Figure \ref{qr}:
\begin{center}
\textcolor{magenta}{\href{https://borterionmusic.bandcamp.com/album/this-silence-made-me-feel-empty}{This silence Made me feel Empty}}
\end{center}

\subsection{Bonus}
To demonstrate the versatility of the algorithm as a musical composition tool, an additional work belonging to the Techno genre is also presented, where the musicalization serves as another instrument. In this piece, it is represented in the form of electric keyboards and acid synthesizers. This second work is titled \textit{I have Made it through this heavy storm}. It can be accessed through the following link: 
\begin{center}
\textcolor{magenta}{\href{https://borterionmusic.bandcamp.com/album/i-have-made-it-through-this-heavy-storm}{I have Made it through this heavy storm}}
\end{center}

\section{Conclusions}
In the first instance, a compositional tool was successfully developed that uses satellite images of clouds as its core, allowing its integration into any desired musical genre. This tool adds polyrhythmic properties that enrich the sonic texture of the final work, adding a unique and complex dimension to the musical composition. The flexibility of the algorithm not only facilitates its adaptation to various musical forms but also provides a versatile platform for the exploration and creation of new artistic expressions based on scientific data.

The research demonstrates the significant potential of integrating satellite data and sonification techniques in algorithmic music production. This approach not only enriches the artistic realm but also opens new avenues for the dissemination of space sciences through music. Such algorithms can be incorporated into audiovisual presentations to create multisensory experiences aimed at scientific outreach. Although it may not be necessary to explain the composition process when presenting the music, it is relevant to mention its origin in clouds as a natural phenomenon, acknowledging these masses of water as co-authors of the work.

This type of project is ideal for contexts where it is necessary to demonstrate the breadth of topics covered by the sciences, showing that they are not limited to the abstract imagination but can also lead to interdisciplinary projects involving artistic expressions. An example of these spaces is university career fairs, where one can appreciate how science and art can converge in a diversity of projects.

Finally, if my sonifications spark the interest of even one person in the sciences, they will have fulfilled their purpose. Thank you for reading.

\bibliography{clouds}

\clearpage
\appendix

\end{document}